# Locations of logistics facilities for e-commerce: a case of the Tokyo Metropolitan Area

Takanori Sakai[a,*], Kohei Santo[a], Shinya Tanaka[a], Tetsuro Hyodo[a]

[a]*Department of Logistics and Information Engineering, Tokyo University of Marine Science and Technology, 2-1-6 Etchujima Koto-ku Tokyo Japan, 135-8533*

**Abstract**

The rapid growth of the e-commerce market creates new dynamics in the logistics landscape, which has been evolving for decades in cities around the world. It is a challenge for businesses and planners to meet the high demand for logistics facilities for e-commerce order fulfillment and goods handling. In the Tokyo Metropolitan Area, mega-scale multi-tenant logistics facilities have been developed in both the port area near the urban center and the periphery of the city, while delivery service providers (DSPs) locate many last-mile delivery stations, varying in number depending on the urban density. We analyze the spatial distribution and location factors of both mega-scale multi-tenant facilities and last-mile delivery facilities. We found that, due to the scarcity of land, newly developed multi-tenant facilities are more likely to be in less accessible places that have high-level development restrictions. The result also indicates the heterogeneity of the distribution of DSPs' facilities, reflecting the heterogeneity in business strategies.

## 1. Introduction

In the last several decades, logistics land use has been changing dramatically in cities around the world. With the innovations in logistics systems and supply chains that have realized fast and reliable goods flows, the main function of logistics facilities has shifted from storage to goods handling and transshipment. Contemporary supply chains require logistics facilities that have a much larger footprint than traditional warehouses. Due to the shortage of available land in cities, many of those logistics facilities have been developed in the suburbs and exurbs. These spatial dynamics of logistics facilities are called "logistics sprawl." Logistics sprawl has been observed in many cities around the world (Aljohani and Thompson, 2016; He et al., 2018), including the Tokyo Metropolitan Area (TMA) (Sakai et al., 2015, 2017). Practitioners and researchers alike are concerned about the inefficiency of the spatial distribution of logistics facilities and the associated externalities of freight transportation (Dablanc and Rakotonarivo, 2010; Sakai et al., 2019).

* Corresponding author. Tel.: +81-3-5245-7461.
*E-mail address:* tsakai2@kaiyodai.ac.jp

The rapid growth of the e-commerce market, which the COVID-19 pandemic accelerates, adds a new dynamism to the spatial distribution of logistics facilities. In Japan, the share of the e-commerce market in the retail sector grew by 6.76% in 2019 and 8.08% in 2020 (Ministry of Economy, Trade, and Industry, Japan, 2021). In the U.S., the country with the highest adoption rate of e-commerce, retail e-commerce sales reached 14.3% of the total retail market in the first quarter of 2022 (U.S. Department of Commerce, 2022). Although the services are convenient for consumers, it is argued that B2C last-mile deliveries are more fragmented than B2B goods flows, and therefore, their negative impacts are concerning (Beckers et al., 2022; Sakai et al., 2022). E-commerce logistics systems have a different design intent from those serving in-store retail businesses. For consumers to receive a variety of goods in a short lead time, fulfillment facilities are often required to be extremely large and close to consumers. Furthermore, to deliver many parcels to households and workplaces, many small 'last-mile facilities' need to be deployed in the neighborhoods of these delivery destinations. Physical spaces serving as origins are required for last-mile deliveries. Moreover, the deliveries from them should be able to adapt to the urban environment (e.g., mobility and parking restrictions). There is a gap between available spaces in cities, which are highly limited and expensive, and the needs of e-commerce-driven logistics land use demand. Although some studies identified that logistics sprawl has not necessarily led to inefficiency in logistics systems and an increase in negative externality (Sakai et al., 2017; Sakai et al., 2019; Kang, 2020a), the resulting excess land use in suburbs and/or exurbs would be more problematic for e-commerce logistics than B2B logistics. Due to such concerns, policy development for "proximity logistics", i.e., logistics facilities in dense and mixed-use urban areas, is identified as an important research subject (Buldeo Rai et al., 2022).

The purpose of this research is to understand the spatial characteristics of logistics facilities that are relevant to the intra-metropolitan legs of e-commerce freight demand. Aiming to obtain valuable insights for logistics land use policies, we analyze the data from the Tokyo Metropolitan Area (TMA), which is the largest metropolitan area in the world. We mainly focus on two types of logistics facilities: multi-tenant (MT) and last-mile (LM) facilities. MT facilities are for-lease facilities that are usually extremely large (typically 30,000 $m^2$ or larger) with multiple-story structures. In the TMA, Prologis developed their first MT facility in 2007, and, as of 2022, they operate around 40 facilities in the TMA. The use of MT facilities is very common for e-commerce vendors (or e-tailers). The regional distribution hubs and fulfillment centers of major vendors that dominate the market, such as Amazon, Rakuten, and Zozotown, are using the MT facilities. On the other hand, the LM facilities are facilities operated by delivery service providers (DSPs), which are typically very small, and function as the origins of the last-mile deliveries fulfilled by DSPs to the end consumers. In Japan, more than 90% of parcel deliveries are handled by three major DSPs (Yamato Transport, Sagawa Express, and Japan Post), and they operate the LM facilities. (However, it is anecdotally known that Amazon Japan increasingly counts on non-major carriers and sole proprietors, for which public statistical data is not available.) Taking into account the local context, this research investigates the spatial distributions and location factors of those two types of e-commence-related logistics facilities.

The rest of the paper is organized as follows: in Section 2, we provide a review of literature focusing on the spatial distribution and location factors of logistics facilities at the metropolitan scale; in Section 3, we discuss the key conditions relevant to logistics facility development, such as urban structure and regulatory policies, as well as the recent development trend of e-commerce-related logistics facilities; in Section 4, we analyze the spatial distribution of different types of logistics facilities, comparing MT and LM facilities with other logistics facilities; in Section 5, we develop location choice models and discuss key location factors for e-commerce-related logistics facilities; finally, Section 6 concludes the research, summarizing the research contributions and the future tasks.

## 2. Literature review

In recent years, the spatial distribution of logistics facilities has been discussed frequently in the context of "logistics sprawl" (Aljohani and Thompson, 2016). In academic articles, the term "logistics sprawl" was first used by Dablanc and Rakotonarivo (2010), and since then, many studies have measured the outward migration of logistics facilities using the average distance of logistics facilities from the mean center (barycenter) or the urban center (e.g., Dablanc et al., 2012; Sakai et al., 2015; Woudsma et al., 2016; Giuliano and Kang, 2018; Heitz et al., 2020; Kang, 2020a). The accumulative knowledge from the studies of various cities around the world indicates that, while exceptions exist, logistics sprawl is a global phenomenon that occurs where the modernization of logistics practices caters to the constraints of urban agglomeration, such as limited space, high land prices, and traffic congestion in urban centers. In most of the past studies of logistics sprawl, the locational difference by functional type of logistics facilities was not considered due to data limitations, and thus, discussions addressing the heterogeneity in the spatial patterns of logistics facilities were limited. Exceptions that take advantage of rich data are Sakai et al. (2020b), which studied the spatial patterns by the origin-destination types of shipments handled (i.e., inter-city and intra-city shipments), and Sakai et al. (2018), which analyzed the patterns by commodity type handled.

With the growing interest in e-commerce, Xiao et al. (2021) attempt to obtain insights on the influence of e-commerce on logistics land-use trends by combining the spatial data of logistics facilities (for which function details are not available) and interviews with stakeholders, including logistics companies, retailers and e-tailers, logistics associations, government departments, logistics real estate developers, and consulting companies. They found a spillover of logistics land demand from Shenzhen to neighboring cities, which is likely to be driven by the shortage of land in Shenzhen and the growth in the e-commerce market. Rodrigue (2020) presents a case study of Amazon’s distribution network in the U.S., highlighting the functional and spatial positions of logistics facilities for e-commerce order fulfillment. The study indicates that a considerable share (about 70%) of the facility footprint is dedicated to e-fulfillment centers, followed by delivery stations (about 14%), which are LM facilities.

Along with the growth of interest in spatial distribution, the location factors of logistics facilities have also been studied more frequently. The studies use the framework of regression and/or discrete choice analysis to measure the effects of accessibility and locational characteristics on location choice (Hagino and Endo, 2007; Woudsma et al., 2008; Durmus and Turk, 2014; Verhetsel et al., 2015; Yuan, 2018; Gingerich and Maoh, 2019; Kang, 2020b; Sakai et al., 2020b). However, these studies also do not investigate the differences in location factors by functions of logistics facilities, except for the size of the facilities. As an exception, Sakai et al. (2020a) developed a location choice model for each function group of logistics facilities in the Paris Region, France. Their study identified heterogeneity in the effects of accessibility (to end-consumers) on logistics facility locations. Also, the study found significant impacts of land use regulations and traditional clusters on logistics facility locations. However, the study uses relatively broad facility categories (only five categories are considered) and therefore the findings as to the locations of logistics facilities that are highly relevant to e-commerce-drive deliveries are not provided). Past studies thus provide only limited insights as to the locations of logistics facilities that handle large volumes of e-commerce-driven shipments, such as MT and LT facilities.

As the above review suggests, the studies of e-commerce-related logistics facilities are still in their infancy and lag behind the rapid growth of the e-commerce market, despite the interest in the topic. This study strives to address this gap by analyzing the spatial distributions and location factors of two major facilities for e-commerce-driven shipments, MT and LM facilities, in the TMA and obtaining policy insights for addressing emerging issues associated with the rise of e-commerce.

## 3. Environment for logistics land use in the TMA

### *3.1. Urban structure*

The study area includes four prefectures: Tokyo, Kanagawa, Saitama, and Chiba, and the southwest part of Ibaraki (15.7 thousand $km^2$), where about 3.8 million people resided as of 2015. The TMA has a monocentric urban structure with the Tokyo Railway Station at the center (Fig.1). Though the demand for land is very high, resulting in high land prices, the industrial zone remains around the center of the metropolitan area. Furthermore, the industrial region facing Tokyo Bay, called the Keihin region, is still one of the largest industrial regions in Japan. The industrial development near the urban center and the coastal area is a legacy of industrial modernization in the late 19th century, when nation-wide modernization occurred. The study area also includes the busiest seaports (Tokyo, Kawasaki, and Yokohama) and airports (Narita and Tokyo) in the country. The most recent infrastructure development that has impacted the urban logistics system in the TMA is the completion of the Ken-O Expressway sections, which function as the 3rd Ring Road. More than half of all sections have been completed during the period 2013–2017, leaving only the 18.5 km section in the east (close to Narita Airport) incomplete, which turned the areas along the road into competitive candidates for large-scale logistics developments. As discussed later, many newly developed logistics facilities are located along the Kei-O Expressway (KOE).

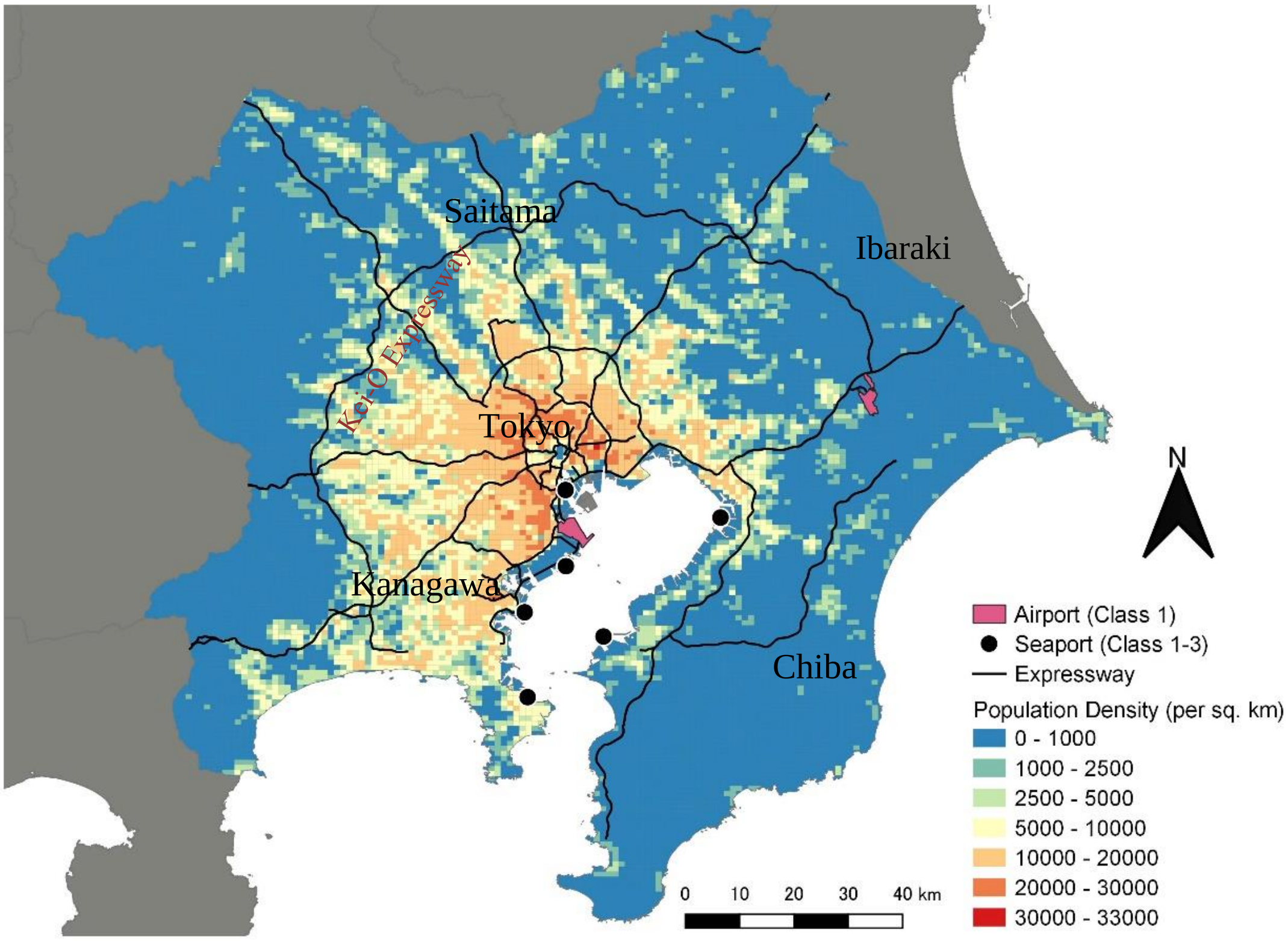

Fig. 1. Population and transportation infrastructure in the study area.

### *3.2. Regulatory policy*

Among urban development policies, zoning regulations and urbanization control are the main tools that navigate the locations of logistics facility development (Fig.2). In the TMA, zoning regulations, which are specified within the Urbanization Promotion Area (UPA), are not highly restrictive for developing logistics facilities as all categories of commercial and industrial land use accept logistics facilities, and even a residential land use category – quasi-residential zone, which is for areas along arterial roads – allows for logistics facilities. As a drawback, the current zoning regulations could generate conflict between residents and logistics operators, which is often triggered by the encroachment of residential development into areas that were historically industrial. Some local municipalities impose policies to control such development (e.g., Sagamihara City, Kanagawa, and Toda City, Saitama).

On the other hand, almost all types of developments, including logistics facilities, are strictly restricted in the Urbanization Control Area (UCA), which functions as greenbelts that prevent urban sprawl. However, with the recent completion of the KOE and the pressing demand for logistics land use development, local municipalities impose policies that ease such restrictions. For example, Saitama prefecture in the north part of the TMA converted a large area of UCA along the KOE (about 4.4 km2) into UPA to attract industrial and logistics developments in the area under the policy named "Industrial Zone Development Policy in the Garden City," effective from 2006 to the present day. Also, at the national level, the program "Act on Advancement of Integration and Streamlining of Distribution Businesses (AAISDB)" enacted in 2005 grants permission for the development of logistics facilities in UCA if such development leads to greater efficiency of logistics systems. It must be noted, however, that AAISDB does not cover MT logistics facilities, where tenants are not fixed.

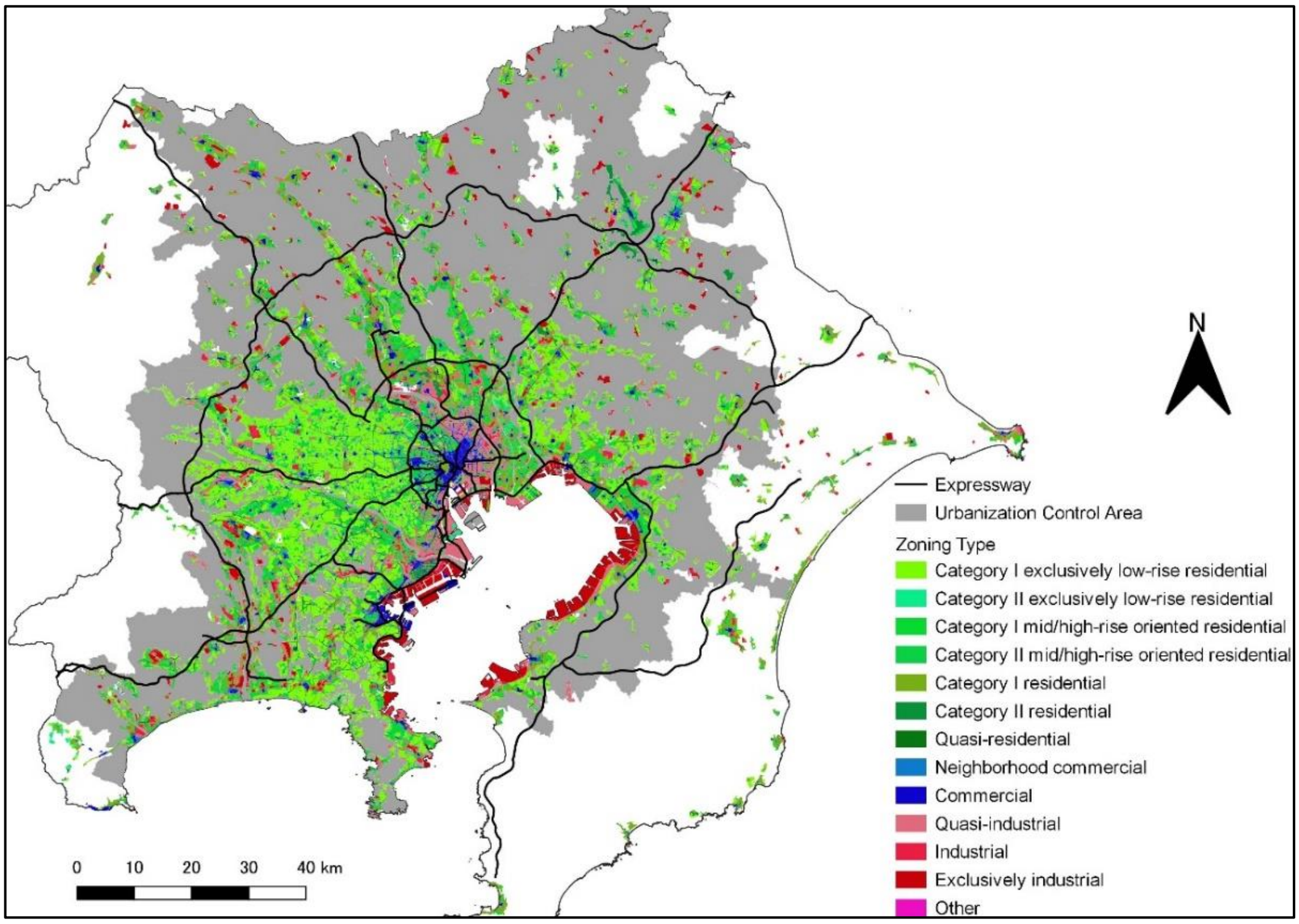

Fig. 2. Zoning regulations.

### *3.3. Development trend*

The supply of logistics facilities has been growing very fast in recent years. According to Japan Logistics Field Institute Inc. (2022), more than 3 million m$^2$ of logistics real estate (mostly MT facilities) were supplied in the Greater Tokyo Area (seven prefectures including the study area) in 2021 alone, which is the greatest annual supply in history. At the national level, about 30% of logistics real estate (on a floor area basis) is used for e-retailing, which is much greater than the share of e-retailing in 2016, about 10%. Another data source, CBRE (2021), indicates that about 150 MT facilities (confirmed existing as of 2021), equivalent to 13 million m$^2$, were added during the period 2011–2021 in the study area. It must be noted that those MT facilities are used not only for fulfillment and cross-docking but also for last-mile deliveries, which take only a small share but are increasing (Logistics Field Institute Inc., 2022). As for the LM facilities, due to the strict enforcement for illegal delivery vehicle parking started in 2006 (due to the amendment of the Road Traffic Act), which is impactful especially in high-density areas, DSPs have adapted their facility scale and transportation mode (truck, van, bicycle, handcards) based on the urban landscape (business district, high-rise residential district, low-rise residential district). It must be noted that, unlike MT facilities, there is no time-series data available for LM facilities, and thus the quantitative evidence of the change in supply (floor area) is lacking.

## 4. Analysis of the spatial distribution of logistics facilities relevant to e-commerce

### *4.1. Data*

To understand the spatial distribution of e-commerce-related logistics facilities, we use the data of logistics facilities in the TMA grouped into five sets, which are summarized in Table 1. The first dataset is large-scale MT logistics facilities with information on floor area and year of establishment, published by CBRE (2021), a global real estate services and investment firm. Fig. 3 shows examples of MT facilities. The data covers 155 MT facilities that are larger than 30,000 m$^2$. All these facilities were developed in 2011 or later. We confirmed, using Google Map (https://www.google.com/maps), this data cover most of or all the MT facilities we intended to analyze. We enriched this data with the coordinates of positions obtained from Google Map. We label this dataset "MT".

Table 1. Logistics facility datasets.

| Label | Sample size | Ave. floor area (m$^2$) | Description |
|---|---|---|---|
| MT | 155 | 88,069 | MT facilities which are larger than 30k m$^2$, developed in 2011 or later. |
| LMY | 988 | n/a | LM facilities operated by Yamato Transport, which exist as of August 2022. The information of size and establishment year are not available. |
| LMS | 290 | n/a | LM facilities operated by Sagawa Transport, which exist as of August 2022. The information of size and establishment year are not available. |
| TMFS-L | 293[1)] | 18,775[2)] | Logistics facilities which are larger than 5k m$^2$ developed during 2000-2013, sampled in the 2013 Tokyo Metropolitan Freight Survey. |
| TMFS-S | 157[1)] | 131[2)] | Logistics facilities which are smaller than 300 m$^2$ developed during 2000-2013, sampled in the 2013 Tokyo Metropolitan Freight Survey. |

Note: 1) With expansion; 2) Considering expansion factors.

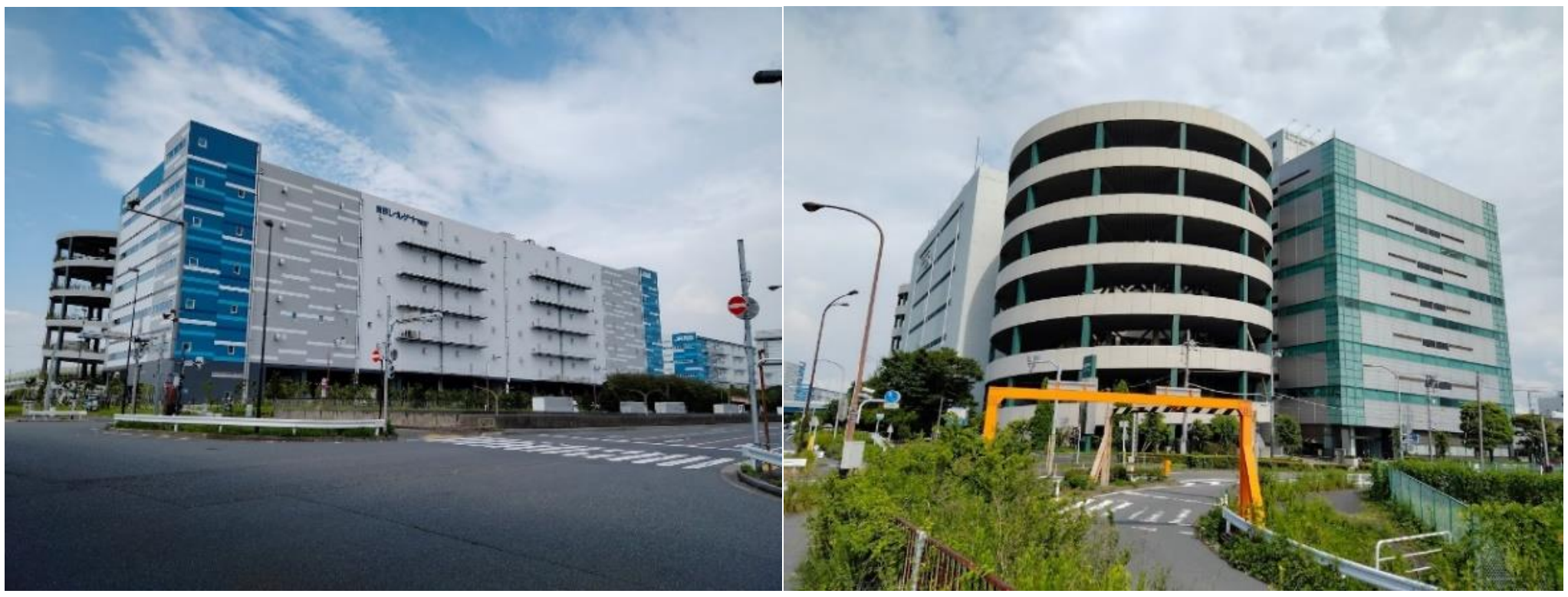

Fig. 3. Examples of MT facilities

The second and third datasets are the locations of delivery stations (i.e., LM facilities) operated by Yamato Transport and Sagawa Express, respectively. We first used two web mapping platforms, Navitime (https://www.navitime.co.jp/) and Google Map, to identify the address and coordinates of the facilities, and then we checked their facility type visually using Google Street View and removed non-LM facilities such as cross-docking facilities where many large goods vehicles park; even if such facilities are used for last-mile deliveries, they are not taken into account. The examples of the LM facilities, where hand carts, cargo bikes, and vans are used for deliveries, are shown in Fig. 4. In many cases, two or more branch names are assigned to a single physical facility; we further removed duplicate addresses and generated a list of unique facilities (988 facilities for Yamato Transport and 290 facilities for Sagawa Express). We did not include the data of Japan Post Service, one of the three major DSPs, as their establishments serve multiple different purposes, including customer service, banking and insurance services, management, and logistics (for postal and package deliveries), and were considered unsuitable for the purpose of the analysis. We label this data for Yamato Transport and Sagawa Express as "LMY" and "LMS," respectively. We did not combine these two sets of data because the two companies do not necessarily locate their LM facilities following the same principles.

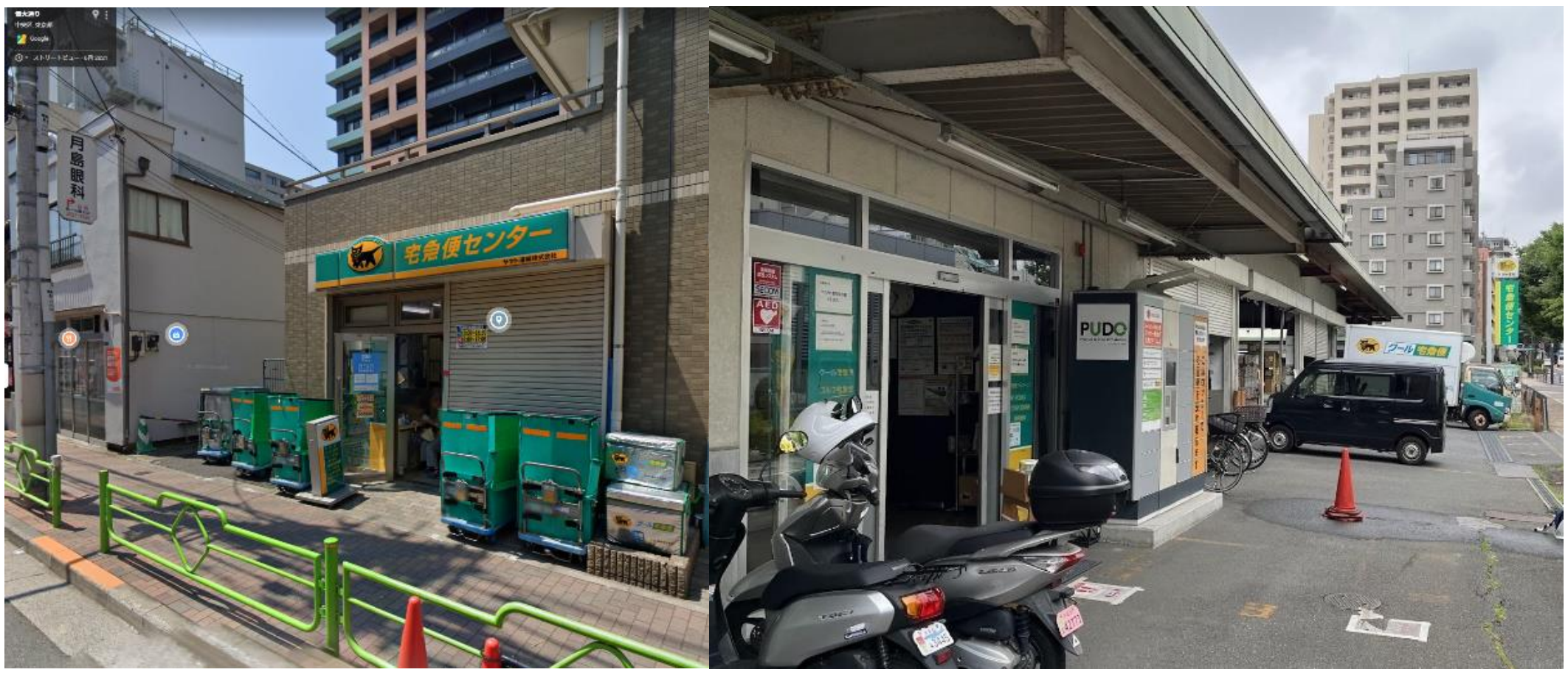

Fig. 4. Examples of LM facilities.

The third dataset is the one from the 2013 Tokyo Metropolitan Freight Survey (TMFS), which was also used in Sakai et al. (2017) and Sakai et al. (2020). The survey was conducted from October 2013 to January 2014 and collected data from 43,131 business establishments, including 4646 logistics facilities. From the dataset, we extracted the data of 293 logistics facility establishments with 5,000 $m^2$ or larger (we label this dataset "TMFS-L") and that of 157 establishments with 300 $m^2$ or smaller (we label this dataset "TMFS-S"), both of which were established from 2000 to 2013. Note that TMFS-L and TMFS-S are sample data and most of these facilities were established before or at the early stage of the growth of the MT facility market in the TMA. We use this data as a reference for the distributions of large- and small-scale general logistics facilities to highlight the uniqueness of spatial distributions of MT and LM, respectively. It is preferred to use the data of even larger facilities from TMFS as a reference for the analysis of MT and smaller facilities as a reference for the analysis of LM. However, further narrowing the range of facility scale significantly reduces the sample sizes for TMFS-L and TMFS-S and thus these data become unusable. For each sample of the TMFS data, an expansion factor is computed by the Transport Planning Commission of the Tokyo Metropolitan Region (TPCTMR), which implemented the survey, based on location type, type of industry, and employment size. We use those expansion factors to correct the sampling bias for the results shown below.

### *4.2. Spatial characteristics of logistics facilities*

For visualizing the spatial distributions of logistics facilities, we use the Kernel Density Estimation (KDE) method with the Gaussian distribution. We use ‘spatstat’ package in R (https://CRAN.R-project.org/package=spatstat), test different bandwidth values to find the one that effectively highlights the characteristics of the spatial distributions, and finally set it to 3 km for all groups. Fig. 5 shows the spatial distributions of MT and TMFS-L. Based on the figures, few large logistics facilities are located within the first ring road, the Central Circular Route (an 8 km radius of the center), even though around 35% of that area is for industrial use. The TMFS-L figure shows the presence of two major clusters facing Tokyo Bay, indicating the intensive use of the coastal area. The figure of MT, on the other hand, shows a higher level of spatial dispersion of clusters than that of TMFS-L. Another difference between MT and TMFS-L is that some MT facilities, specifically the facilities on the east side, are located away from the expressway sections. However, the comparison of the average distance from the urban center indicates that, as a whole, MT facilities are located closer to the urban center (67.9 km for MT and 72.2 km for TMFS-L), even though they are much larger than the facilities in TMFS-L. The formation of multiple clusters prevents the dispersion of facilities within suburbs and exurbs, which, as a result, works against the trend of logistics sprawl at the metropolitan level. This trend could be explained in part by the policies of local municipalities, which have strict controls over development in UCA.

As for LM/small-sized facilities, the comparison between LMY, LMS, and TMFS-S is shown in Fig. 6. They have similar spatial distributions, following the urban density (i.e., the density of population and business establishments). The levels of concentration are, however, quite different between LMY and LMS. The company of LMY has engaged in parcel delivery services since the late 1970s, and their facilities cover urbanized areas entirely. On the other hand, the company of LMS was originally more focused on B2B deliveries and started parcel delivery services more recently, in 1998. It is likely that the differences in the past business strategies played a role in the difference, resulting in the relatively high concentration of LMS in the central area (especially the business district).

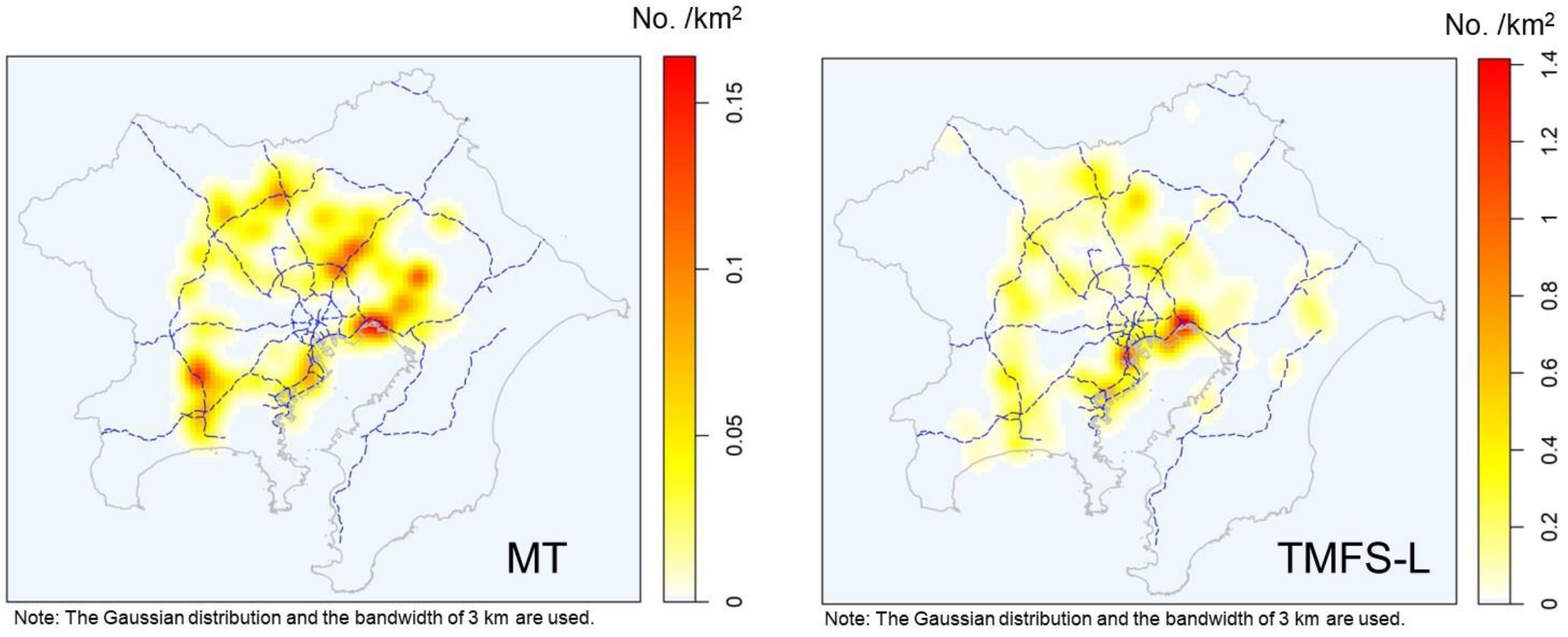

Fig. 5. The spatial distribution of logistics facilities (left: MT, right: TMFS-L).

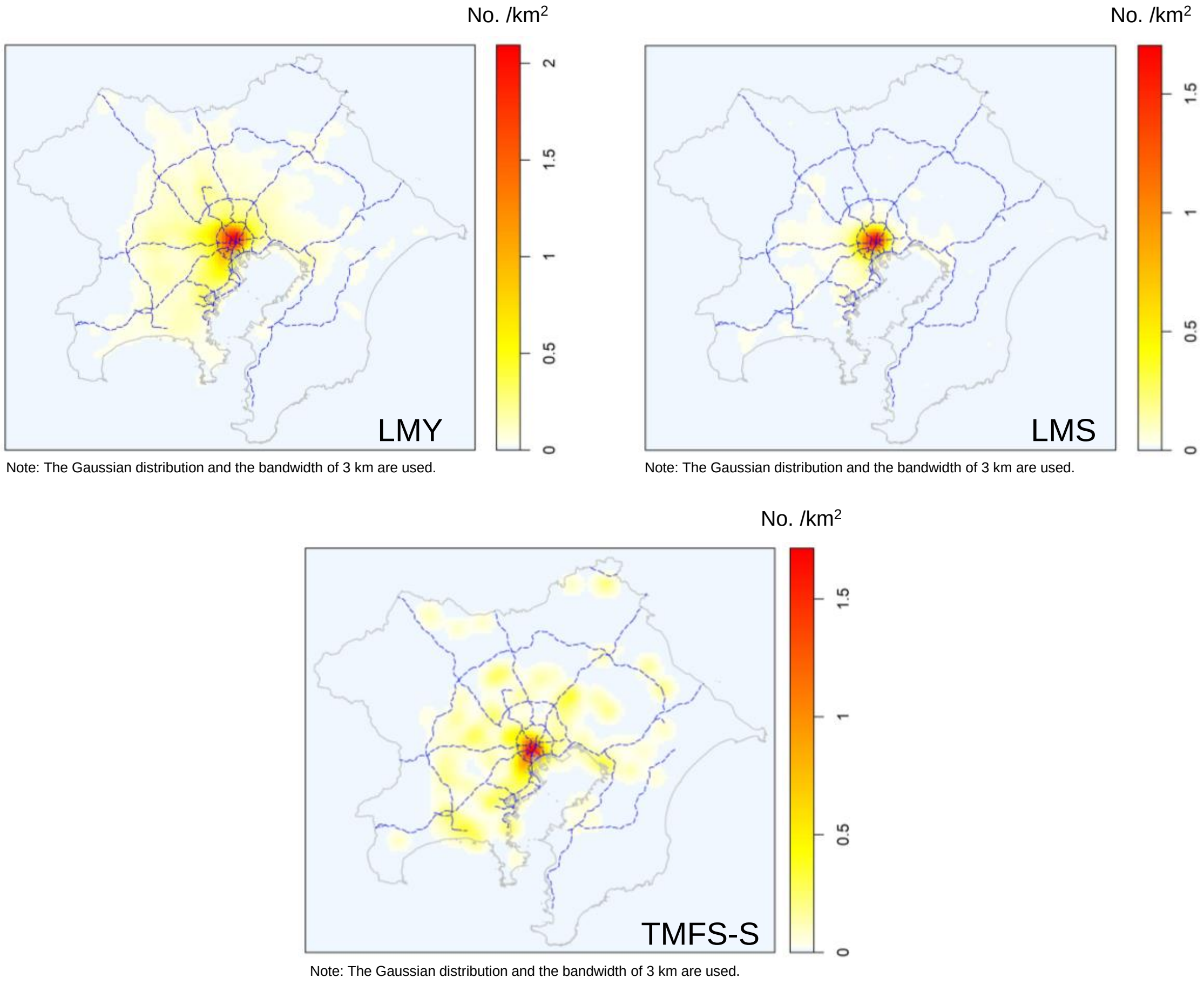

Fig. 6. The spatial distribution of logistics facilities (upper left: LMY, upper right: LMS, bottom: TMFS-S).

## 5. Analysis of location factors

### *5.1. Location choice modeling*

To evaluate the effects of location factors for MT and LM facilities, we estimate multinomial logit (MNL) models for the five facility datasets (MT, TMFS-L, LMY, LMS, TMFS-S) discussed in the previous section. The choice set is 13,682 1 km-by-1 km squares in the study area. This is the highest resolution for which the official statistics such as the Economic Census and the National Census are available to us. The locations that do not include any developable land (i.e., all areas are wasteland, forest, and/or water) are removed from the choice set. The variables that characterize each location are summarized in Table 2. It must be noted that we use the most recent data available, which means the data does not necessarily match the establishment years of facilities; the same approach is adapted by the past studies like Kang (2020b) and Sakai et al. (2020a).

Table 2. Independent variables considered.

| Variable | Description | Expected sign of effect |
|---|---|---|
| log(no. of employees) | Log-transformed employment size based on the 2016 Economic Census. | + |
| log(distance to expressway) | Log-transformed distance to the nearest expressway interchange (as of 2021) in kilometers. | - |
| log(avg. distance from the nearest facilities in the same group) | Log-transformed average distance from the nearest X[1)] facilities in the same group (MT, TMFS-L, LMY, LMS, or TMFS-S) in kilometers. | - |
| log(population density) | Log-transformed population density in thousand per $km^2$ based on the 2015 National Census. | - (MT, TMFS-L, TMFS-S), + (LMY, LMS) |
| log(avg. land price) | Log-transformed average land price in million yen per $m^2$ based on the 2019 Official Land Price Data. | - |
| Share of residential zone<br>Share of commercial zone<br>Share of quasi-industrial zone<br>Share of industrial zone<br>Share of exclusively industrial zone<br>Share of urbanization control area<br>Share of other (miscellaneous land use and non-urban planning area) | Share of land within each polygon that is zoned for residential, commercial, etc. (Share of residential zone is considered as the base category.) | |

Note: 1) X is determined so that it will be roughly 3% of sample size; MT:5, TMFS-L:10; LMY: 30; LMS:10; TMFS-S: 10.

The utility $U_{i,l}$ of location $l$ for logistics facility $i$ is specified as follows:

$$U_{i,l} = V_l + \varepsilon_{i,l} \tag{1}$$

$$V_l = \boldsymbol{\beta} \cdot \boldsymbol{X_l} \tag{2}$$

where:
$V_l$: Deterministic component of the utility.
$\varepsilon_{i,l}$: Identically and independently Gumbel distributed random component (scale =1, location=0).

$\boldsymbol{X_l}$: A vector of the characteristics of location $l$.
$\boldsymbol{\beta}$: A vector of the parameters to be estimated.

The probability $P_l$ for a logistics facility to be located at location $l$ is:

$$P_l = \frac{exp\ (V_l + lnA_l)}{\sum_{l' \in L} exp\ (V_{l'} + lnA_{l'})} \tag{3}$$

where:
$A_l$: available area for development at location $l$

The summary statistics of independent variables are shown in Table 3. All variables are normalized (mean=0, SD=1) before the model estimation.

Table 3. Independent variables considered (before log-transformation and normalization).

| Variable | Mean | Median | SD | Min. | Max |
|---|---|---|---|---|---|
| No. of employees (thousand) | 1.3 | 0.2 | 6.0 | 0.0 | 232.2 |
| Distance to expressway (km) | 10.5 | 8.4 | 7.6 | 1.0 | 59.7 |
| Avg. distance from the nearest facilities in the same group (km) | | | | | |
| -MT | 20.0 | 15.3 | 14.9 | 0.4 | 66.5 |
| -TMFS-L | 17.4 | 13.4 | 12.5 | 0.1 | 62.7 |
| -LMY | 16.7 | 15.0 | 9.9 | 0.7 | 49.8 |
| -LMS | 23.3 | 22.5 | 12.5 | 0.0 | 60.1 |
| -TMFS-S | 20.4 | 19.0 | 11.3 | 0.7 | 58.2 |
| Population density (thousand per $km^2$) | 3.2 | 0.7 | 5.0 | 0.0 | 36.4 |
| Avg. land price (million yen per $m^2$) | 0.12 | 0.03 | 0.47 | 0.00 | 21.32 |
| Share of commercial zone (ratio) | 0.02 | 0.00 | 0.08 | 0.00 | 1.00 |
| Share of quasi-industrial zone (ratio) | 0.02 | 0.00 | 0.08 | 0.00 | 1.00 |
| Share of industrial zone (ratio) | 0.01 | 0.00 | 0.05 | 0.00 | 0.86 |
| Share of exclusively industrial zone (ratio) | 0.02 | 0.00 | 0.09 | 0.00 | 1.00 |
| Share of urbanization control area (ratio) | 0.28 | 0.03 | 0.36 | 0.00 | 1.00 |
| Share of other (ratio) | 0.15 | 0.00 | 0.28 | 0.00 | 1.00 |

### *5.2. Estimated models*

The estimated models for MT and TMFS-L are shown in Table 4. Two models were estimated for each of them: with and without "*log(avg. distance from the nearest facilities of the same group)*" (Model1 and Model2). The independent variables for which the estimated coefficients show the opposite signs to those expected were removed to eliminate unintended multicollinearity problems. Through the estimation process that tests various combinations of variables, "*log(avg. land price)*" often shows the opposite sign and/or very low t-values. Usually, higher land prices are associated with higher rents (e.g., rents are higher in port areas than inland areas). From the perspective of the owners, higher revenue is expected for the higher-value locations despite the higher land costs. In the study of US metropolitan areas conducted by de Oliveria et al. (2022), it is shown that land or rent value is not significantly related to logistics sprawl. Our results do not contradict their findings. The comparison of Model 1 indicates that, compared with TMFS-L, MT is less sensitive to accessibility indicators (*no.*

*of employees and distance to expressway*), less separated from local residents (*population density*), but more influenced by zoning regulations (see the estimated coefficients of the shares of industrial zones), as well as local municipalities' attitudes to acceptance (indicated by the strong effect of *share of urbanization control area* ($\beta = 0.475$). One of the reasons could be that MT facilities require larger spaces than the facilities included in TMFS-L. This clearly indicates the need for reasonable logistics land use policies targeting extremely large and increasingly common MT facilities. Further, the comparison of Model 2 indicates that MT are more likely to cluster (inferred from *avg. distance from the nearest facilities in the same group*) than logistics facilities in general. The trend of "logistics clusterization" has been observed in various contexts and in different countries (Sakai et al., 2023). MT facilities have a relatively high tendency to cluster, potentially because the locations with high market value as logistics real estate are the same among logistics real estate companies. The tenants of MT facilities are not fixed, and therefore the locations need to be attractive to a wide range of different demands.

Next, the estimated models for LMY, LMS, and TMFS-S are shown in Table 5. Again, two models are estimated for each dataset. From the comparison of Model 1, we can observe the heterogeneity in location factors. *No. of employees* has the strongest effect on LMS ($\beta = 2.460$) while *population density* has the strongest effect on LMY ($\beta = 1.654$). TMFS-S shows the negative coefficient of *log(population density)* ($\beta = -0.204$), indicating those facilities, which are typically for B2B shipments, tend to avoid populated areas, while the LM facilities (LMY and LMS) show the opposite tendency. Generally speaking, the effects of zoning regulations are much weaker for small facilities than for MT and TMFS-L. As for Model 2, the result shows that both LMY and LMS (in addition to TMFS-S) show the tendency to cluster; *log(avg. distance from the nearest facilities in the same group)* has negative coefficients ($\beta = -0.497$ for LMY and $\beta = -0.158$ for LMS). This is surprising, as it is typically considered that LM facilities in the same company are located apart from one another. This is likely because DSPs attempt to serve high delivery demand locations not by locating larger facilities but by locating more facilities of similar (small) sizes, which are easier to find.

Table 4. Estimated location choice models (MT and TMFS-L).

| Variables | MT | | | | TMFS-L | | | |
|---|---|---|---|---|---|---|---|---|
| | Model 1 | | Model 2 | | Model 1 | | Model 2 | |
| | Coef. | t-val. | Coef. | t-val. | Coef. | t-val. | Coef. | t-val. |
| log(no. of employees) | 0.923 | 4.6 | 0.712 | 3.2 | 1.175 | 14.1 | 1.017 | 11.6 |
| log(distance to expressway) | -0.353 | -3.7 | -0.077 | -0.8 | -0.617 | -15.2 | -0.222 | -4.7 |
| log(avg. distance from the nearest facilities in the same group) | - | - | -1.447 | -14.9 | - | - | -0.965 | -30.8 |
| log(population density) | -0.223 | -2.4 | -0.284 | -3.0 | -0.402 | -12.1 | -0.384 | -11.8 |
| log(avg. land price) | - | - | -0.194 | -1.5 | - | - | - | - |
| Share of commercial zone | -0.304 | -2.4 | -0.139 | -1.2 | -0.211 | -6.0 | -0.164 | -4.7 |
| Share of quasi-industrial zone | 0.206 | 4.7 | 0.164 | 3.5 | 0.150 | 8.9 | 0.062 | 3.4 |
| Share of industrial zone | 0.217 | 7.6 | 0.180 | 5.8 | 0.184 | 15.0 | 0.197 | 16.2 |
| Share of exclusively industrial zone | 0.288 | 6.8 | 0.173 | 3.5 | 0.201 | 12.1 | 0.175 | 10.1 |
| Share of urbanization control area | 0.475 | 3.8 | 0.315 | 2.3 | 0.294 | 5.7 | 0.342 | 6.6 |
| Share of other | -5.997 | -1.7 | -6.487 | -1.6 | -0.146 | -1.6 | 0.110 | 1.2 |
| L(0) | -1476 | | -1476 | | -8625 | | -8625 | |
| L($\hat{\beta}$) | -1274 | | -1162 | | -7349 | | -6974 | |
| $\rho^2$ | 0.137 | | 0.213 | | 0.148 | | 0.191 | |
| $\bar{\rho}^2$ | 0.131 | | 0.206 | | 0.147 | | 0.190 | |

Table 5. Estimated location choice models (LMY, LMS and TMFS-S).

| Variables | LMY | | | | LMS | | | | TMFS-S | | | |
|---|---|---|---|---|---|---|---|---|---|---|---|---|
| | Model 1 | | Model 2 | | Model 1 | | Model 2 | | Model 1 | | Model 2 | |
| | Coef. | t-val. | Coef. | t-val. | Coef. | t-val. | Coef. | t-val. | Coef. | t-val. | Coef. | t-val. |
| log(no. of employees) | 1.654 | 15.9 | 1.288 | 11.9 | 2.460 | 10.5 | 2.425 | 10.3 | 1.804 | 16.7 | 1.455 | 12.2 |
| log(distance to expressway) | -0.133 | -3.1 | - | - | -0.245 | -2.5 | -0.193 | -1.9 | -0.196 | -4.2 | -0.091 | -1.8 |
| log(avg. distance from the nearest facilities of same group) | - | - | -0.497 | -8.8 | - | - | -0.158 | -6.1 | - | - | -0.690 | -12.4 |
| log(population density) | 0.800 | 7.1 | 0.527 | 4.9 | 0.452 | 2.7 | 0.283 | 1.8 | -0.134 | -2.4 | -0.204 | -3.7 |
| log(avg. land price) | | | | | -0.031 | -0.4 | -0.094 | -1.1 | | | -0.298 | -5.4 |
| Share of commercial zone | 0.054 | 3.3 | 0.032 | 1.9 | 0.113 | 4.2 | 0.069 | 2.4 | -0.044 | -2.2 | -0.047 | -2.2 |
| Share of quasi-industrial zone | 0.025 | 1.3 | 0.015 | 0.8 | -0.014 | -0.3 | -0.023 | -0.6 | 0.062 | 3.3 | 0.035 | 1.8 |
| Share of industrial zone | 0.018 | 0.8 | 0.029 | 1.4 | -0.001 | 0.0 | -0.006 | -0.1 | 0.016 | 0.8 | 0.001 | 0.1 |
| Share of exclusively industrial zone | -0.065 | -1.3 | -0.030 | -0.6 | 0.090 | 1.2 | 0.057 | 0.8 | -0.066 | -2.0 | -0.094 | -2.6 |
| Share of urbanization control area | 0.461 | 6.7 | 0.403 | 6.0 | 0.128 | 0.7 | 0.044 | 0.2 | 0.101 | 1.7 | 0.049 | 0.8 |
| Share of other | 0.233 | 2.5 | 0.292 | 3.1 | 0.295 | 1.2 | 0.238 | 1.0 | -0.222 | -2.0 | -0.141 | -1.3 |
| L(0) | -9381 | | -9381 | | -2762 | | -2762 | | -7296 | | -7296 | |
| L($\hat{\beta}$) | -8029 | | -7992 | | -1905 | | -1890 | | -6465 | | -6393 | |
| $\rho^2$ | 0.144 | | 0.148 | | 0.310 | | 0.316 | | 0.114 | | 0.124 | |
| $\bar{\rho}^2$ | 0.143 | | 0.147 | | 0.307 | | 0.312 | | 0.113 | | 0.122 | |

## 6. Conclusion

We focused on the two types of logistics facilities which are highly relevant to e-commerce deliveries in the urban environment, MT and LM facilities, and analyzed their spatial distributions and location factors in the TMA. By comparing these e-commerce-related facilities with other logistics facilities, we aimed to understand how the growth of the e-commerce market could affect the spatial dynamics of logistics land use. The result reveals that their location choice mechanisms are distinct from those of other types of logistics facilities. MT facilities do not necessarily contribute to the outward migration of logistics facilities. However, they tend to be in less accessible locations than large facilities developed in the past. Zoning regulations are more influential and, therefore, important for MT facilities. Also, the developments of the MT facilities have sought exceptional development permissions in highly controlled areas (i.e., UCA), which potentially were justified with the recent completion of the KOE (the 3rd Ring Road) sections, although those exceptional arrangements might not be allowed in the long-term. In short, with the increasing use of MT facilities, the role of public policy for logistics land use is more important than before. LM facilities, on the other hand, are less constrained by zoning and tend to be located closer to residential areas and business offices, adapting distribution modes to local road infrastructure. Nonetheless, their distribution patterns differ by company, and the company's historical business practices appear to have an impact on the current spatial distributions, despite the fact that the supply of space for LM facilities may not be constrained in the TMA.

To the best of our knowledge, this is the only research that focuses on and unveils the spatial characteristics of MT and LM facilities, which are highly relevant to e-commerce at the metropolitan scale, except for the one focusing on the U.S. Amazon (Rodrigue, 2020). Future studies focusing on other cities, together with the findings from this study, are expected to provide a more generalizable (and global) picture of the dynamics of logistics land use created by the growth in the e-commerce market. Furthermore, more work is required to gain knowledge about the relationship between the spatial distribution of e-commerce-related facilities and the impact of associated traffic. Such knowledge has been limited even for logistics facilities in general, despite the accumulation of studies on logistics sprawl in the past (Sakai et al., 2019). In particular, the traffic impacts associated with extremely large MT facilities are worthy of investigation. Those facilities generate a large number of truck trips, and thus the relationship between their locations and associated negative externalities is important to understand for policy development.

## Acknowledgements

We would like to thank the Transport Planning Commission of the Tokyo Metropolitan Region for sharing the data for this research.